\documentclass[twocolumn,showpacs,floats,prl,aps,unsortedaddress,superscriptaddress]{revtex4-1}

\usepackage[dvips]{graphicx}
\usepackage{amsfonts}
\usepackage{amsmath}
\usepackage{amssymb}
\usepackage{exscale}
\usepackage{color}	
\usepackage{braket}	
\usepackage{multirow}	

\begin{document}
\title{Momentum-resolved evolution of the Kondo lattice into `hidden-order' in URu$_2$Si$_2$}

\author{F.~L.~Boariu}
\thanks{F.L.B. and C.B. contributed equally to this work}
\affiliation{Lehrstuhl f\"ur Experimentelle Physik VII, Universit\"at W\"urzburg, Am Hubland, D-97074 W\"urzburg, Germany}
\author{C.~Bareille}
\thanks{F.L.B. and C.B. contributed equally to this work}
\affiliation{CSNSM, Universit\'e Paris-Sud and CNRS/IN2P3, B\^atiments 104 et 108, 91405 Orsay cedex, France}
\author{H.~Schwab}
\affiliation{Lehrstuhl f\"ur Experimentelle Physik VII, Universit\"at W\"urzburg, Am Hubland, D-97074 W\"urzburg, Germany}
\author{A.~Nuber}
\affiliation{Lehrstuhl f\"ur Experimentelle Physik VII, Universit\"at W\"urzburg, Am Hubland, D-97074 W\"urzburg, Germany}
\author{P.~Lejay}
\affiliation{Institut N\'eel, CNRS/UJF, B.P. 166, 38042 Grenoble Cedex 9, France}
\author{T.~Durakiewicz}
\affiliation{MPA-CMMS, Los Alamos National Laboratory, Los Alamos, New Mexico 87544, USA}
\author{F.~Reinert}
\affiliation{Lehrstuhl f\"ur Experimentelle Physik VII, Universit\"at W\"urzburg, Am Hubland, D-97074 W\"urzburg, Germany}
\affiliation{Karlsruher Institut f\"ur Technologie (KIT), Gemeinschaftslabor f\"ur Nanoanalythik, D-76021 Karlsruhe, Germany}
\author{A.~F.~Santander-Syro}
\email{andres.santander@csnsm.in2p3.fr}
\affiliation{CSNSM, Universit\'e Paris-Sud and CNRS/IN2P3, B\^atiments 104 et 108, 91405 Orsay cedex, France}

\begin{abstract}
	We study, using high-resolution angle-resolved photoemission spectroscopy,
	the evolution of the electronic structure in URu$_2$Si$_2$ at
	the $\Gamma$, $Z$ and $X$ high-symmetry points from the high-temperature
	Kondo-screened regime to the low-temperature `hidden-order' (HO) state.
	At all temperatures and symmetry points, we find structures resulting from the
	interaction between heavy and light bands, related to the Kondo lattice formation.
	At the $X$ point, we directly measure a hybridization gap of 11~meV already open
	at temperatures above the ordered phase. 
	Strikingly, we find that while the HO induces pronounced changes at $\Gamma$ and $Z$,
	the hybridization gap at $X$ does not change, 
	indicating that the hidden-order parameter is anisotropic.
	Furthermore, at the $\Gamma$ and $Z$ points, we observe the opening of a gap in momentum 
	in the HO state, and show that the associated electronic structure results
	from the hybridization of a light electron band with the Kondo-lattice
	bands characterizing the paramagnetic state.     
\end{abstract}
\maketitle

The heavy-fermion URu$_{\bf 2}$Si$_{\bf 2}$ presents
a second-order phase transition at $T_{HO} = 17.5$~K
to a `hidden order' (HO) state of yet unknown order parameter~\cite{palstra85, maple86, schlabitz86}.  
The 27-year quest for an understanding of this transition
has triggered an extensive research~\cite{broholm87,schoenes87,dawson89,
  mason91,broholm91,santini94, buyers94,escudero94,amitsuka99,chandra02,chandra02pmagn,bourdarot03,wiebe04,behnia05,wiebe07,
  elgazzar09,janik09,santander09,haule09,haule10,yoshida10,schmidt10,aynajian10, oppeneer10,
  oppeneer11,kawasaki11,dakovski11,dubi11,haraldsen11,pepin11,riseborough12,chandra12,mydosh11}.
The properties of this material are determined by the dual
`itinerant-localized' character of the uranium $5f$ electrons,
with Kondo screening developing below $T \sim 70$~K, as inferred from
transport data~\cite{palstra85, maple86}.
Earlier angle-resolved photoemission spectroscopy (ARPES) experiments
indicated the presence, in the paramagnetic (PM)
state, of an $f$-like feature at the Fermi level ($E_F$) near the $X$
point~\cite{denlinger00,denlinger01},
while optical conductivity data showed that a Drude peak forms below 75~K, 
consistent with metallic behaviour~\cite{bonn88,levallois11,nagel11,hall12}.
Thus, a crucial aspect of the HO is that it emerges on a pre-formed Kondo lattice.  
Indeed, recent high-resolution ARPES and scanning tunneling microscopy experiments 
demonstrated that itinerant heavy quasiparticles
participate in the Fermi-surface instability at the HO transition~\cite{santander09,yoshida10,schmidt10,aynajian10}.
However, to date, there is no momentum-resolved picture spanning several high-symmetry points 
showing how the electronic structure evolves from the Kondo-screened regime to the HO state.

In this work, we demonstrate the existence of distinct heavy-fermion features 
at the $X$, $\Gamma$, and $Z$ points of URu$_2$Si$_2$ 
up to temperatures close to the onset of Kondo screening.
We show that these structures result from the hybridization between heavy and light bands,
and can be thus linked to the formation of the Kondo lattice. 
In particular, at the $X$ point, we directly observe a hybridization gap of $\sim 11$~meV 
fully open at $T > T_{HO}$.
We find that the HO transition shifts the Kondo-lattice structures 
at the $\Gamma$ and $Z$ points well below $E_F$,
while leaving unchanged the hybridization gap at $X$,
explicitly showing that the order parameter 
does not affect equally all the bands near $E_F$.
Additionally, we observe that in the HO state, 
the heavy-fermion bands at $\Gamma$ and $Z$ become gapped in momentum at $E_F$. 
We provide a phenomenological model to describe the electronic structure at $X$, $\Gamma$ and $Z$ 
and its evolution from the PM Kondo-screened state to the HO state. 
In particular, we show that a light electron band (LEB), 
interacting with the two bands from the Kondo lattice, 
is an essential ingredient to understand the observations below $T_{HO}$ at $\Gamma$ and $Z$.  

\begin{figure}[b]
  \begin{center}
   	  \includegraphics[clip, width=8cm]{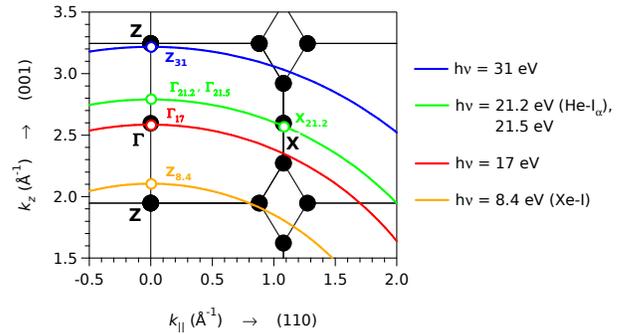}
  \end{center}
  \caption{\label{fig:BZ} \footnotesize{ 
  		Body-centered tetragonal Brillouin zone (black lines) and ARPES measurement arcs (color lines)
  		for photon energies of 8.4~eV (Xe-I), 17~eV, 21.2~eV (He-I$\alpha$), $21.5$~eV and 31~eV. 
  		Open circles show the measurement points discussed in the main text.
  		The index of each point refers to the photon energy in eV.
  		The arcs correspond to a model of a free-electron final state 
  		with an inner potential $V_0 = 13$~eV~\cite{denlinger00}.
  		} 
  	}
\end{figure}
\begin{figure*}[t]
  \begin{center}
    \includegraphics[clip, angle=-90,width=15cm]{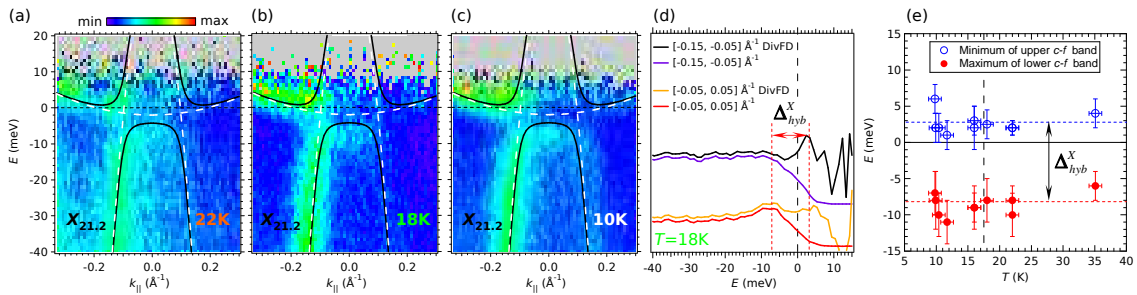}
  \end{center}
  \caption{
    \label{fig:Xpt} \footnotesize{
    	(a-c)~Energy-momentum ARPES intensity maps at the $X$ point of URu$_2$Si$_2$,
        using He-I$_{\alpha}$ photons, at 22~K, 18~K and 10~K, respectively.
        The data have been normalized to the FD distribution of a metallic reference 
        at the same temperature and in electrical contact with the sample, measured under identical
        conditions~\cite{santander09}.
        Intensity differences between left and right image halves are attributed to matrix elements
        changing at $X$ when going across neighboring Brillouin zones.
    	The dashed white lines and solid black lines represent the original and hybridized bands used to fit the data. 
    	(d)~Spectra at 18~K integrated over the maximum of the $\Pi$-shaped band (red line),
    	then divided by FD (DivFD, orange line), and integrated over the minimum of the upper hybridized structure
    	(violet and black lines). The peaks corresponding to the lower and upper 
    	parts of the hybrid structure, and their gap $\Delta_{hyb}^{X} \approx 11$~meV, are clearly observed.
    	Note that the fall-off at $E_F$ of the raw data is much larger than the resolution,
    	indicating the presence of the HEB, as revealed by the division by FD.
        (e) Experimental values of the maximum of the $\Pi$-shaped band (red circles)
        and the minimum of the upper hybridized structure (blue circles) as a function of temperature,
        measured as shown in (d). 
        }  
        }
\end{figure*}

The ARPES experiments were performed with Scienta R4000 detectors 
at W\"urzburg University, using monochromatized He-I$_\alpha$ ($h\nu = 21.2$~eV, resolution 5.18~meV) 
and Xe-I ($h\nu = 8.4$~eV, resolution $\sim 4$~meV) photons from an MBS T-1 multigas discharge lamp, 
and at the UE112-PGM-1b ($1^3$) beamline of the Helmholtz Zentrum Berlin (HZB-BESSY II)  
using horizontally polarized light at $h\nu = 17$,~$21.5$~and~$31$~eV (resolution 3~meV).
Measurements at different $h\nu$ correspond to different values of $k_z$ along $(001)$~\cite{Hufner-Book-2003},
as shown in Fig.~\ref{fig:BZ}.
The samples were cleaved {\it in-situ} along the $(001)$ axis at 10~K (W\"urzburg) and 1~K
(BESSY), and measured along the $(110)$ (or $k_{\parallel}$) direction. 
The pressure was below $5\times10^{-11}$~Torr at BESSY and when using the Xe lamp, separated
from the measurement chamber by a MgF$_2$ window, and of $5\times10^{-10}$~Torr when using the He lamp.
We checked that the superconducting transition at $1.2$~K has no measurable effect on the spectra at 1~K.

We discuss first the data at $X$, whose structure, as we will see, 
can be straightforwardly described in terms of a Kondo hybridization.  
Figures~\ref{fig:Xpt}(a-c) present the ARPES spectra at the $X_{21.2}$ point
at 22~K, 18~K and 10~K, respectively.
The data are essentially identical.
Below $E_F$, one observes a $\Pi$-shaped band, whose flat maximum lies at $E \approx -8$~meV.
Furthermore, division by a Fermi-Dirac (FD) distribution of appropriate effective temperature~\cite{santander09}
reveals the dispersing wings of a heavy electron band (HEB) occurring right above $E_F$.
This type of structure is the hallmark of a Kondo hybridization between
a light hole band (LHB) and a HEB~\cite{misra08, klein11, yang11}.
In particular, as shown in figure~\ref{fig:Xpt}(d), one distinctly observes 
a large hybridization gap $\Delta_{hyb}^{X} \approx 11$~meV {\it already open at} $T=18$~K.
Additional measurements, summarized in figure~\ref{fig:Xpt}~(e), show that this gap
is temperature independent up to $T~\sim 2 T_{HO}$.
In fact, the data at $X$ can be fitted by a standard hybridization model~\cite{misra08} 
between a LHB of mass $\sim -0.9 m_e$ ($m_e$ is the free-electron mass)
and a HEB of mass $\sim 50-70 m_e$ interacting through a potential $V_{he}^{X} \sim 11$~meV.
The original LHB and HEB, and the resulting ``upper" and ``lower" hybridized bands,
are represented by the white dashed and solid black lines in Figs.~\ref{fig:Xpt}(a-c). 
Thus, our data at $X$ provide a direct momentum-resolved imaging 
of a Kondo hybridization gap of 11~meV in URu$_2$Si$_2$, and demonstrate that such a gap
opens well above $T_{HO}$, consistent with the carriers' scattering rate abruptly decreasing
below the same energy scale at $T \lesssim 60-90$~K observed in early optical~\cite{bonn88}
and recent ultrafast reflectivity~\cite{liu11} measurements.

\begin{figure*}
  \begin{center}
  	  \includegraphics[clip, angle=-90,width=15cm]{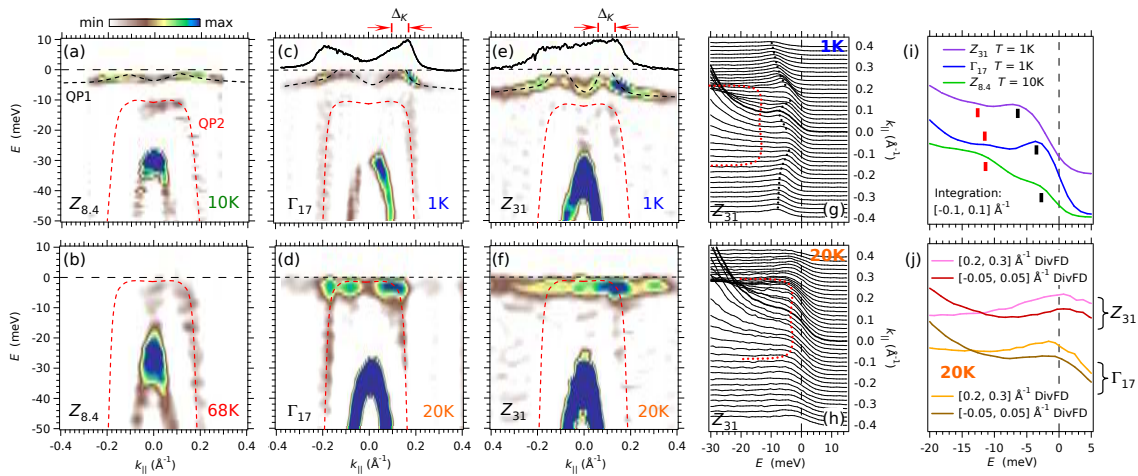}
  \end{center}
  \caption{\label{fig:qp1-qp2-ZGZ} \footnotesize{
  	  (a,~b)~Second derivative of ARPES data at the $Z_{8.4}$ point, at 10~K and 68~K respectively.
      (c-f)~Second derivative of ARPES data at $\Gamma_{17}$ and $Z_{31}$, at 1~K and 20~K.
  	  (g,~h)~Raw EDCs at $Z_{31}$, at 1~K and 20~K.
  	  (i)~Spectra in the HO state integrated around $k_{\parallel} = 0$ at $Z_{8.4}$, $\Gamma_{17}$ and $Z_{31}$.
  	  (j)~Spectra in the PM state, divided by FD, integrated around $k_{\parallel} = 0$ 
  	  and $k_{\parallel} = 0.25$~\AA$^{-1}$ at $\Gamma_{17}$ and $Z_{31}$.
  	  In panels (c,~e), the black solid curves show the MDCs integrated over 5~meV around $E_F$.
  	  A gap in momentum $\Delta_{k} \approx 0.08 \pm 0.01$~\AA$^{-1}$ is indicated by the red arrows.
  	  This gap is also evident from the raw data in panel (g). 
  	  It decreases as temperature raises, and is unresolved at 10~K in panel (a).
      In all panels, the black and red dashed lines or vertical bars are guides to the eye for QP1 and QP2, respectively,
      and the measurement direction is $(110)$. 
      } 
   }
\end{figure*}
We now discuss the evolution of the electronic structure across the HO transition 
at the $\Gamma$ and $Z$ points.
Figures~\ref{fig:qp1-qp2-ZGZ}(a,~b) show the electronic structure at $Z_{8.4}$
in the HO (10~K) and PM (68~K) states
(the raw data and details of the second derivative calculations are presented in the Supplemental Material).
The intense surface state below $-30$~meV and a LHB 
parallel to it were described previously~\cite{santander09,boariu10,yoshida10,yoshida11}.
Furthermore, the data at 10~K in Fig.~\ref{fig:qp1-qp2-ZGZ}(a) show a heavy M-shaped quasiparticle band 
dispersing down to $E = -3$~meV at $k_{\parallel} = 0$ (hereafter QP1, black dashed lines), 
and a second, $\Pi$-shaped band (QP2, red dashed lines), similar to the one observed at $X$, 
with a flat maximum at $E = -10$~meV.
The flat part of QP2 was observed in previous laser-ARPES studies of the $Z$ point~\cite{yoshida10,yoshida11,yoshida12}.
A crucial novel aspect of our data is the visible onset of dispersion of QP2: 
following the flat region around $k_{\parallel} = 0$, 
at momenta larger than $\sim 0.15$~\AA$^{-1}$, QP2 merges with the LHB mentioned above,
forming the $\Pi$-shaped structure that is gapped with respect to $E_F$.
Also new in our data is that, as shown in Fig.~\ref{fig:qp1-qp2-ZGZ}(b), 
the $\Pi$-shaped structure exists at temperatures as high as 68~K, 
close to the onset of Kondo screening,
while previous ARPES studies at the $Z$ point claimed that above $T_{HO}$ 
all features disappeared or were not detectable~\cite{yoshida10, yoshida11, yoshida12}.
However, in contrast to the HO state, at 68~K the binding energy of QP2 is now $\approx E_F$, 
and QP1 is not detected anymore --either because it shifted above $E_F$ or because it merged with QP2. 
Previous reports have shown that, in the HO state, both QP1 and QP2 shift towards $E_F$
as temperatures rises~\cite{yoshida12}. The data of figures~\ref{fig:qp1-qp2-ZGZ}(a~b),
plus data discussed next confirm this picture,
and demonstrate that at all temperatures above $T_{HO}$ and up to 68~K, 
one observes only the peak of QP2 around $E_F$, its binding energy remaining essentially temperature-independent. 

Figures~\ref{fig:qp1-qp2-ZGZ}(c,~d) show the electronic structure at 
the $\Gamma_{17}$ point at 1~K and 20~K. The corresponding data at $Z_{31}$ are presented in 
figures~\ref{fig:qp1-qp2-ZGZ}(e,~f).
Figures~\ref{fig:qp1-qp2-ZGZ}(g,~h) display the raw energy-distribution curves (EDCs) at $Z_{31}$.
Figure~\ref{fig:qp1-qp2-ZGZ}(i) presents data in the HO state integrated around $k_{\parallel} = 0$
at $Z_{8.4}$, $\Gamma_{17}$ and $Z_{31}$.
Similarly, figure~\ref{fig:qp1-qp2-ZGZ}(j) shows data in the PM state,
divided by the appropriate FD distribution, integrated around $k_{\parallel} = 0$ 
and $k_{\parallel} = 0.25$~\AA$^{-1}$ at $\Gamma_{17}$ and $Z_{31}$.
All these figures show that, in the HO state, QP1 and QP2 exist both at $Z$ and $\Gamma$. 
This demonstrates that QP2 is a general feature of the electronic structure along the $(001)$ direction.
Later on, we will show that QP1 and QP2 can be understood on the common framework
of the evolution of the Kondo lattice across the HO transition.    
As seen from figure~\ref{fig:qp1-qp2-ZGZ}(i), 
at 1~K the energies of QP1 and QP2 at $k_{\parallel} = 0$ are
systematically lower than at 10~K. 
These temperature-induced energy shifts of both QP1 and QP2
indicate that both structures are related to the bulk physics of the HO transition~\cite{yoshida12}.
More important, the high-resolution measurements at 1~K, Figs.~\ref{fig:qp1-qp2-ZGZ}(c,~e,~g)
(see also the Supplemental Material), distinctly show that, at $\Gamma$ and $Z$,
the M-shaped band becomes gapped in momentum:
the tips of the M lie above $E_F$, and the dispersion cuts through $E_F$ at two different Fermi momenta, 
$k_F^{inner} \approx \pm 0.06$~\AA$^{-1}$ and $k_F^{outer} \approx \pm 0.14$~\AA$^{-1}$.
On the other hand, at $T > T_{HO}$, Figs.~\ref{fig:qp1-qp2-ZGZ}(b,~d,~f,~h,~j) show that 
QP1 and QP2 have shifted at or near $E_F$ for the three values of $k_z$.
Consequently, the momentum gap at the tips of the M closes.
In particular, at 20~K in $\Gamma_{17}$ and $Z_{31}$, figures~\ref{fig:qp1-qp2-ZGZ}(d,~h,~j), 
one still distinguishes traces of the high-momenta wings of QP1's M-like dispersion.
However, these are now very close to $E_F$,
and are significantly broadened by temperature and by increased scattering to other Fermi momenta
that become available as the HO gap closes 
--similar to the well-known case of quasiparticles in superconducting cuprates~\cite{kaminski00}. 
Thus, in the PM phase at $\Gamma$ and $Z$, it becomes difficult to assess whether a gap between QP1 and QP2
is still present. 

An important outcome of our observations is that, contrary to $\Gamma$ and $Z$, 
at $X$ the gap structure is {\it not} affected by the HO transition.  
This is consistent with transport and optical measurements, which suggest that the HO parameter
is {\it anisotropic} along the Fermi surface~\cite{maple86,behnia05,escudero94,bonn88, hall12}.

Note that, while the data at $\Gamma$ and $Z$ are more complex, they evoke in many aspects the physics encountered at $X$.
Thus, based on the two-band hybridization at $X$, we now suggest a toy model for the spectra near $\Gamma$ and $Z$.
Our goal is to capture the ingredients that appear essential to describe, at those two points, 
the evolution of the electronic structure from the Kondo-lattice regime into the HO state.
For definiteness, we concentrate on the data at $Z_{31}$.
Figure~\ref{fig:3band}(a) shows the second derivative of the ARPES data at $Z_{31}$ in the PM (20~K) phase 
after being normalized by the FD distribution (raw data in the Supplemental Material).
This puts in evidence a HEB dispersing close to $E_F$, as already described in Fig.~\ref{fig:qp1-qp2-ZGZ}(h).
Therefore, in the PM state, the electronic structures at $X$ and $Z$ 
display both a HEB and a LHB meeting near $E_F$, although at $Z$ we cannot measure directly a hybridization gap. 
However, from the data at the $X$ point (figure~\ref{fig:Xpt}), we know that at 20~K the system
{\it is} in a coherent Kondo state. Thus, it is fair to expect that the HEB and LHB observed at 20~K at $Z$
are also hybridized with a potential similar to the one at $X$ 
--even if, in what follows, this hypothesis is not essential.

\begin{figure}[t]
  \begin{center}
  	  \includegraphics[clip, width=8cm]{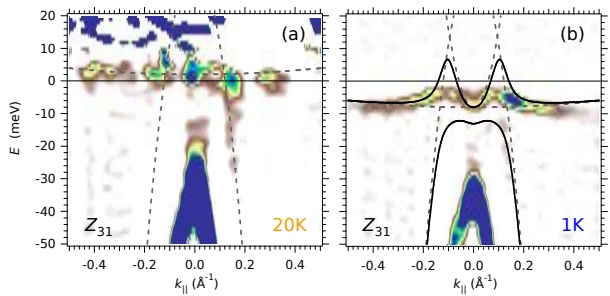}
  \end{center}
  \caption{\label{fig:3band} \footnotesize{(Color online) 
  	  (a,~b) Second derivatives of ARPES data at $Z_{31}$ in the PM (20~K) and HO (1~K) phases,
  	  corresponding to the data in figures~\ref{fig:qp1-qp2-ZGZ}(e,~f) and (g,~h).
  	  The data at 20~K were normalized by the FD distribution of a metallic reference before taking derivatives.
  	  The toy-model's ``original" and hybridized bands are represented by the dashed and solid lines, respectively.
  	  In panel (b), the upper part of the hybridized structure between the LEB and LHB
  	  lies out of the figure scale.
  	  } 
  	  }
\end{figure}

On the other hand, from the data at 1~K in Fig.~\ref{fig:qp1-qp2-ZGZ}(g),
reproduced for clarity in figure~\ref{fig:3band}(b), we note that in the HO state
two additional ingredients are needed to reproduce the peculiar M-shaped dispersion of QP1: 
a strong renormalization (down-shift in energy) of the HEB, 
to account for the heavy high-momenta wings of QP1,
and the introduction of a LEB, to account for the light electron-like dispersion near $k_{\parallel} = 0$.
This last band interacts with the two previously discussed LHB and HEB. As a result, one obtains
the $\Pi$-shaped QP2 below $E \approx -12$~meV and the M-shaped QP1 above $E \approx -7$~meV. 
The best fit is obtained with a LHB 
of mass $-1.6 m_e$ and top energy $35$~meV
hybridizing with a doublet, essentially degenerate at $Z$, 
composed of the HEB (mass $\gtrsim 500 m_e$) and the LEB of mass similar to LHB,
through a hybridization potential $V \approx 11$~meV. 
This potential agrees with the one directly observed at $X$,
reinforcing our expectation for the PM state at $Z$ discussed above.

The main, robust insight from the model above is that, to understand the M-shaped dispersion of QP1 
in the HO state, the hybridization of two bands is {\it not} enough: 
besides the hybrid structure formed by the LHB and the HEB as in the $X$ point, 
one needs a third LEB interacting with the previous two. 
Note also that the interaction with the LEB repels QP2's upper plateau, 
explaining why, below $T_{HO}$, QP1 and QP2 have similar temperature-induced shifts~\cite{yoshida12}.  
Of course, this simple three-band model is limited: in the PM state, we cannot determine accurately the energy of the HEB,
we cannot directly observe a hybridization gap with the LHB,  
and we cannot decide whether the LEB is present slightly above $E_F$, 
because fine details of the unoccupied states cannot be inferred from our data.
Similarly, the model does not reproduce the positive curvature of the QP1 wings at high momenta, 
possibly indicating that a more realistic tight-binding dispersion should be used for the HEB. 

Conservation of particles requires that, in the PM state, the LEB be already present below $E_F$,
possibly at a different region in momentum.
Where this band comes from, and why the HEB drops, are open questions.
One possibility is band nesting or folding~\cite{haule09,elgazzar09,dubi11,riseborough12, chandra12}. 
We note, however, that in standard nesting or folding, 
the energy of the folded band does not shift gradually with temperature~\cite{santander-SCCO-2011},
contrary to the observations, and only the gap between the original and folded bands changes. 

Our results demonstrate that the HO transition is intimately related to the 
Kondo lattice of heavy fermions in URu$_2$Si$_2$, that we directly observe,
including the hybridization gap, up to temperatures {\it well above} $T_{HO}$. 
Furthermore, our data explicitly show that the Fermi-surface instability~\cite{santander09} 
induced by the HO on the Kondo lattice affects differently the electronic structure at various high-symmetry points,
opening a gap in momentum at $\Gamma$ and $Z$.
Regardless of the mechanism behind the HO transition, 
this gap in momentum implies the existence of a gap in energy between the two bands being separated,
that should occur at $E_F$ at other places in reciprocal space.
Our model indicates that such a gap is $\sim 10$~meV, in agreement
with transport experiments~\cite{maple86, bonn88}.  
Crucially, our data analysis strongly suggests that the HO transition is 
related to the interaction between the lattice of heavy fermions {\it and} a band of light electrons,
thus opening gaps in the electronic structure near $E_F$.

We thank C.~P\'epin, M.-A.~M\'easson, S.~Burdin, P.~Chandra and P.~Coleman
for fruitful discussions,  
the HZB for travel funding during the alloted synchrotron radiation beamtime,
and E. Rienks and S. Thirupathaiah for their invaluable help during beamtime.
The work at W\"urzburg is supported by the Deutsche Forschungsgemeinschaft through FOR1162.
T.~D. was supported by the U.S. DOE.
A.~F.~S.-S. acknowledges support from the Institut Universitaire de France. 

\section{Supplemental Material}

\subsection{Kondo hybridization gap around $X_{21.2}$}
\begin{figure*}
  \begin{center}
   	  \includegraphics[clip, width=12cm]{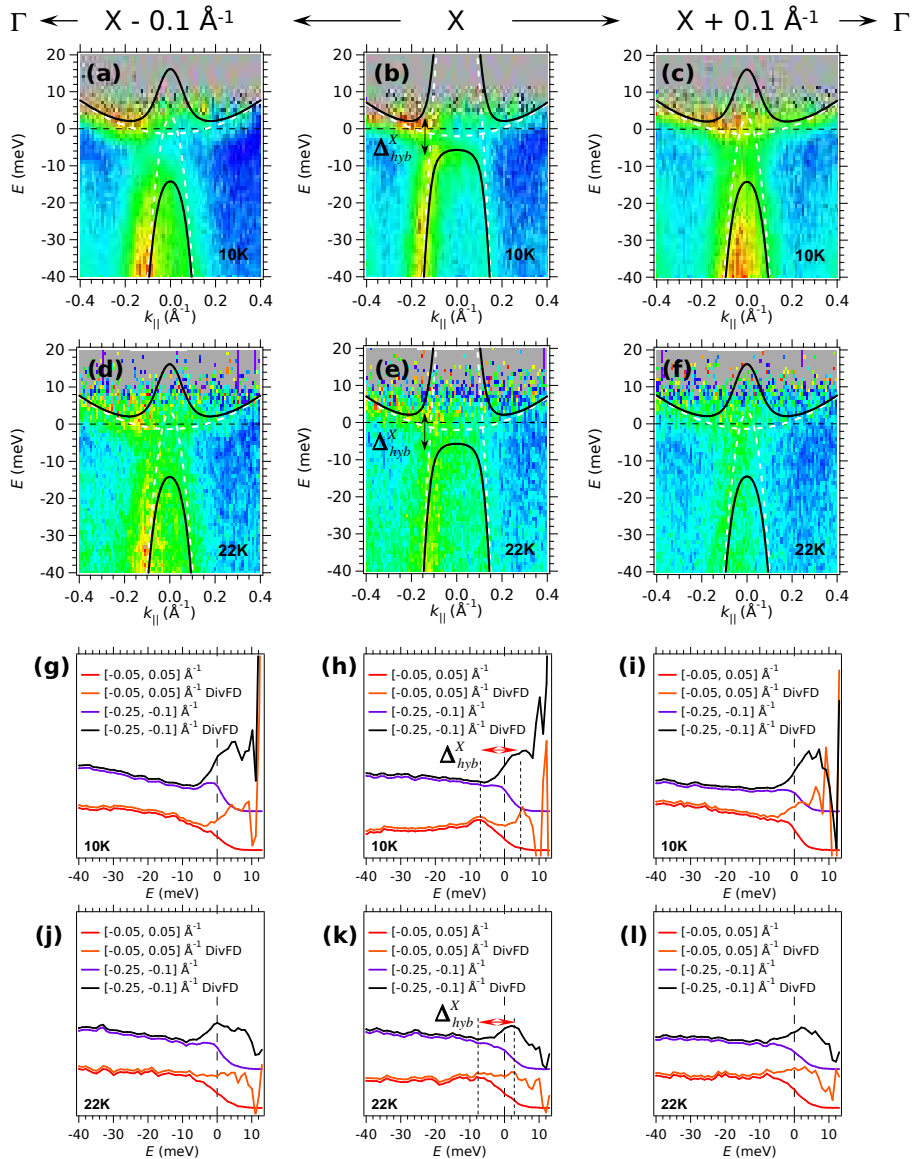}
  \end{center}
  \caption{\label{fig:AroundX} \footnotesize{
  	  	ARPES data at $X$ (central column panels) and at $X \pm 0.1$~\AA$^{-1}$ 
  	  	(right and left columns, respectively) along the $(110)$ direction
  	  	of the body-centered tetragonal Brillouin zone of URu$_2$Si$_2$.
  	  	(a-c) Energy-momentum intensity maps at 10~K, in the HO state.
  	  	(d-f) Corresponding maps at 22~K, in the PM state.
  	  	The data have been normalized to the FD distribution of a metallic reference,
  	  	at the same temperature and in electrical contact with the sample, measured under identical
        conditions~\cite{santander09}.
        Overlaid on the ARPES maps are the original (dashed white lines) and hybridized (solid black lines)
    	bands used to fit the data (described in the text). 
    	(g-i) Spectra at 10~K integrated over the maximum of the $\Pi$-shaped band (red line),
    	then divided by FD (DivFD, orange line), and integrated over the minimum of the upper hybridized structure
    	(violet and black lines). The peaks correspond to the lower and upper 
    	parts of the hybrid structure.  
        (j-l) Same as (g-i) at 22~K.
      } 
      }
\end{figure*}

Figure~\ref{fig:AroundX} presents a wider set of data around the $X$ point, both in the HO (10~K) and PM (22~K)
states, that wholly demonstrate the presence of a hybridization gap already at high temperatures.
The panels in the central column of this figure refer to data exactly at $X$, while the right and left columns
refer to data at $X \pm 0.1$~\AA$^{-1}$, respectively, along the $\Gamma - X - \Gamma$ line in the body-centered
tetragonal Brillouin zone.  At all momenta, at both temperatures, the data show two structures clearly separated in energy:
one heavy electron band (HEB) at or slightly above $E_F$, and one light hole band (LHB) 
with a flat plateau at about $E=-8$~meV occurring exactly at $X$. 
The energy separation between these two bands can be quantified using the integrated
data in panels (g-i), for the data at 10~K, and (j-l) for the data at 22~K. In these integrated spectra,
the peaks correspond to the lower and upper parts of the hybrid structure. Their separation increases with momentum,
simply due to the rapid dispersion of the light hole band. 
The hybridization gap $\Delta_{hyb}^{X}$ 
is experimentally defined as the minimum energy separation between the two hybrid structures, 
which takes place for the spectra going through $X$, as shown in figures~\ref{fig:AroundX}(b,~e,~h,~k).

We find that all the bands at $X$ and $X \pm 0.1$~\AA$^{-1}$ can be fitted with just two bands, one HEB and one LHB,
of essentially circular cross-section in the momentum plane, 
interacting through a momentum-independent potential $V_{he}^{X} \sim 11$~meV, similar to the experimental gap.
Thus, the same bands that fit the data at $X$, figures~\ref{fig:AroundX}(b~e), immediately fit the data
at $X \pm 0.1$~\AA$^{-1}$, just dispersing them by $\pm 0.1$~\AA$^{-1}$.
More important, as with the data at 18~K in the main text, 
the data at 22~K, in particular figures~\ref{fig:AroundX}(e)~and~(k),
also show unambiguously that the hybridization gap is already fully open at $T>T_{HO}$. 

\subsection{Raw data at $Z_{8.4}$}
\begin{figure}
  \begin{center}
   	  \includegraphics[clip, width=8cm]{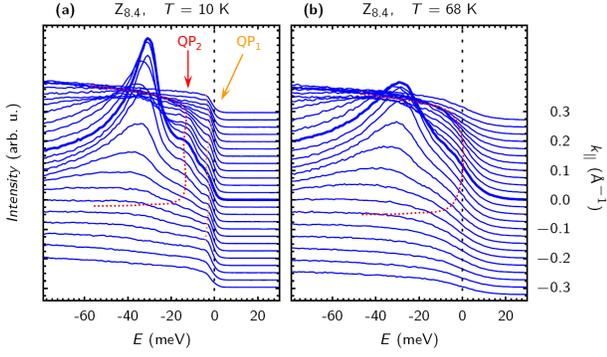}
  \end{center}
  \caption{\label{fig:z-tdep} \footnotesize{ 
      (a,~b) EDC stacks of ARPES data at $Z_{8.4}$, using Xe-I photons, at 10~K and 68~K respectively.
      The measurement direction is $(110)$. 
      The orange and red dotted lines are guides to the eye for QP1 and QP2, respectively.
      The values of $k_{\parallel}$ shown at the right-hand side scale
      refer to the baseline (zero-intensity level well above $E_F$) of the spectra.
      The bold EDCs correspond to $k_{\parallel} = 0$
      } 
      }
\end{figure}
  
Figures~\ref{fig:z-tdep}(a,~b) present the raw data for the electronic structure at $Z_{8.4}$,
{\it i.e.}, using Xe-I photons,
as momentum-resolved stacks of energy distribution curves (EDCs) in the HO (10~K) and PM (68~K) states.
The surface state below $-30$~meV~\cite{santander09,boariu10}, 
and the M-shaped QP1 and $\Pi$-shaped QP2, described in the main text,
are all unambiguosly observed in these figures. 
Note in particular, in panel (a), the clear onset of dispersion of QP2 at momenta larger than $\sim 0.15$~\AA$^{-1}$,
corresponding to hybridization of a heavy electron band with a light hole band, as described in the main text.
Note also, from panel (b), how the flat, non-dispersive part of QP2 has effectively shifted to $\approx E_F$ at 68~K.

\subsection{Raw data at the $\Gamma_{17}$, $\Gamma_{21.5}$ and $Z_{31}$ points}
\begin{figure}
  \begin{center}
  	  \includegraphics[clip, width=8cm]{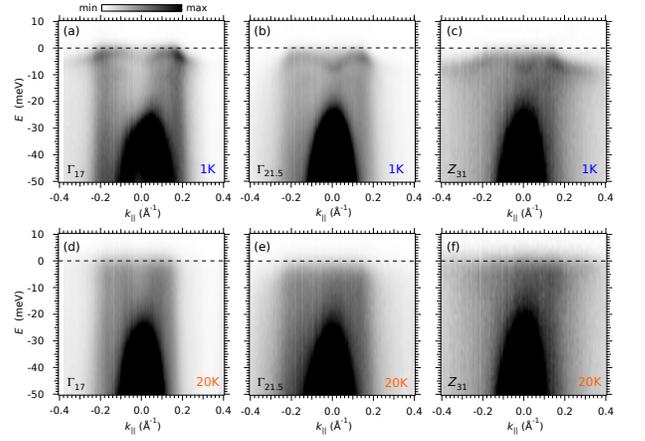}
  \end{center}
  \caption{\label{fig:G17-G21-Z31-Color} \footnotesize{
  	  (a-c) Energy-momentum ARPES intensity maps in the HO state (1~K) at the $\Gamma_{17}$, $\Gamma_{21.5}$ and $Z_{31}$
  	  points, respectively.
  	  (d-f) Corresponding data in the PM state (20~K).
      } 
   }
\end{figure}

\begin{figure}
  \begin{center}
  	  \includegraphics[clip, width=8cm]{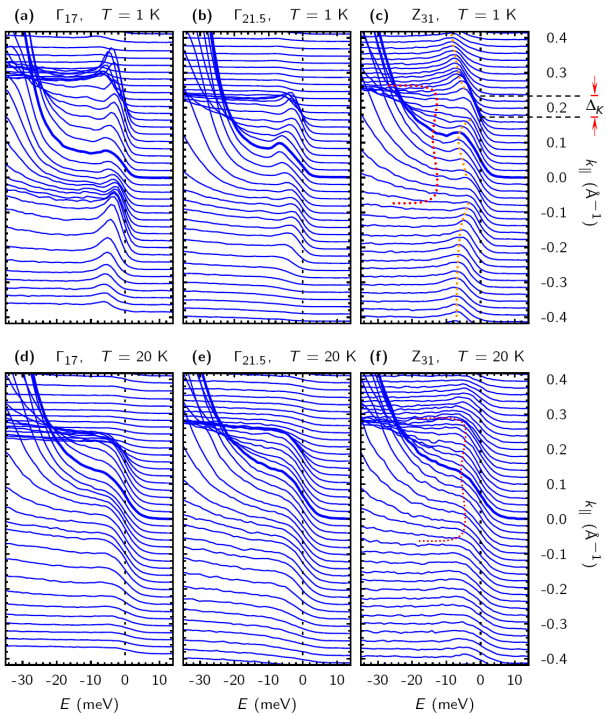}
  \end{center}
  \caption{\label{fig:qp2-002} \footnotesize{ 
  	  (a-c) EDC stacks of ARPES data in the HO state (1~K) at the $\Gamma_{17}$, $\Gamma_{21.5}$ and $Z_{31}$
  	  points, respectively.
  	  (d-f) Corresponding stacks in the PM state (20~K).
  	  The orange and red dotted lines in panels (c) and (f) (point $Z_{31}$) 
  	  are guides to the eye for QP1 and QP2, respectively.
  	  At 1~K, the gap in momentum on QP1 is distinctly observed.
  	  Note also that, at 20~K, the wings of the heavy electron band associated to QP1 are still observed,
  	  but are now located very near to or at $E_F$.
      The values of $k_{\parallel}$, referred to the baseline of the spectra, are shown at the right-hand side scale.
      The bold EDCs denote $k_{\parallel} = 0$.
      All the EDCs in this figure were extracted from the intensity maps shown in figure~\ref{fig:G17-G21-Z31-Color}.  
      } 
   }
\end{figure}

Figure~\ref{fig:G17-G21-Z31-Color} shows the raw energy-momentum ARPES intensity maps 
at 1~K (top row panels) and 20~K (bottom row) for three different photon energies corresponding to the 
$\Gamma_{17}$ (left column), $\Gamma_{21.5}$ (middle column) and $Z_{31}$ (right column) points.

The fine features of the spectra are best observed by representing the data as EDC stacks,
as shown in figure~\ref{fig:qp2-002}.
At 1~K, Figs.~\ref{fig:qp2-002}(a-c), both QP1 and QP2 are observed at the three values of $k_z$.

Note, from figures~\ref{fig:G17-G21-Z31-Color}(a-c) and~\ref{fig:qp2-002}(a-c), 
that QP1 clearly disperses through $E_F$, yielding a gap in momentum
$\Delta_{k} \approx 0.08 \pm 0.01$~\AA$^{-1}$, described in the main text.

At $T > T_{HO}$, Figs.~\ref{fig:qp2-002}(d-f), the $\Pi$-shaped structure ascribed to QP2
shifts to $E \approx E_F$ for the three values of $k_z$.
Moreover, the 20~K data at $\Gamma_{17}$ and $Z_{31}$ also show that  
the upper part of QP2's $\Pi$-shaped structure has actually an electron-like character,
in agreement with the hybridization model presented in the main text. 
Additionally, as can be seen in Fig.~\ref{fig:qp2-002}(f), at 20~K the high-momenta wings of QP1
still give a finite spectral weight at $E_F$, indicating that the heavy band forming QP1,
which was located well below $E_F$ in the HO phase [Fig.~\ref{fig:qp2-002}(c)],
has now shifted to energies near $E_F$, thus closing the momentum gap observed in the HO phase. 

\subsection{Raw data at $Z_{31}$ and band-hybridization model}
\begin{figure}
  \begin{center}
  	  \includegraphics[clip, width=8cm]{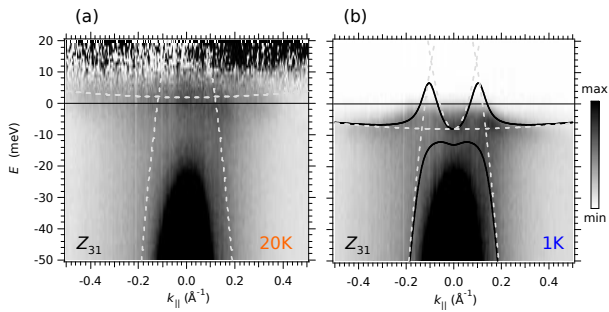}
  \end{center}
  \caption{\label{fig:3band} \footnotesize{
  	  (a,~b) ARPES intensity maps at $Z_{31}$ in the PM (20~K) and HO (1~K) phases, respectively,
  	  corresponding to the EDC stack from Figs.~\ref{fig:qp2-002}(c,~f).
  	  Dotted and solid lines represent, accordingly, the ``original" and hybridized bands
  	  of the scenario accounting for the dispersions of QP1 and QP2 in the HO phase (see main text).  
  	  The upper part of the hybridized structure between the LEB and LHB
  	  lies out of the figure scale.
  	  } 
  	  }
\end{figure}

Figures~\ref{fig:3band}(a,~b) show, over a larger energy range, the raw ARPES intensity maps at $Z_{31}$ 
in the PM (20~K) and HO (1~K) phases. 
The data at 20~K, normalized by the Fermi-Dirac distribution,
shows the wings of a heavy electron band dispersing close to $E_F$.
The proposed non-hybridized bands are represented by the dashed lines in figures~\ref{fig:3band}(a,~b),
and the resulting hybridized bands by the black solid lines.

\subsection{Procedure of second-derivative rendering}
The raw photoemission intensity maps were convoluted with a two-dimensional Gaussian
of widths $\sigma_{E}=3$~meV and $\sigma_{k}=0.06$~\AA$^{-1}$ for temperatures
below 10~K, and $\sigma_{E}=5$~meV and $\sigma_{k}=0.08$~\AA$^{-1}$ for $T > 10$~K.  
Second derivatives along the $E_{B}$ and $k_{\parallel}$ axes were
normalized to the maximum intensity of the surface state peak, then averaged.
Only negative intensity values, which represent peak maxima in the original data, are shown in the main text. 


\end{document}